# Effect of nuclear structure on particle production in relativistic heavy-ion collisions using a multiphase transport model

P. Sinha,[1] V. Bairathi,[2,*] K. Gopal,[1] C. Jena,[1] and S. Kabana[2]

[1]*Department of Physics, Indian Institute of Science Education and Research (IISER) Tirupati, Tirupati 517507, Andhra Pradesh, India*

[2]*Instituto de Alta Investigación, Universidad de Tarapacá, Casilla 7D Arica 1000000, Chile*

(Received 2 June 2023; accepted 11 August 2023; published 25 August 2023)

We report a study of transverse momentum ($p_T$) spectra for $\pi^\pm$, $K^\pm$, $p$, and $\bar{p}$ in isobar, ${}^{96}_{44}$Ru + ${}^{96}_{44}$Ru and ${}^{96}_{40}$Zr + ${}^{96}_{40}$Zr, collisions at $\sqrt{s_{NN}} = 200$ GeV using a multiphase transport model. Particle yields ($dN/dy$), average transverse momenta ($\langle p_T \rangle$), and particle ratios are reported in various collision systems with different parametrizations of the Woods-Saxon distribution. We observed a maximum difference of 5% in the particle yields in peripheral collisions when we included a quadrupole and octupole deformation and a nuclear size difference between the isobars. The $\pi^-/\pi^+$ ratio is smaller in Ru + Ru collisions compared with Zr + Zr collisions indicating an effect of isospin due to difference in number of protons and neutrons between the two nuclei. The $K^-/K^+$ ratio is same in both the systems indicating the dominance of the pair production mechanism in the kaon production. The $\bar{p}/p$ ratio is further smaller in Ru + Ru collisions than Zr + Zr collisions, indicating the effect of baryon stopping in addition to the isospin effect. A system size dependence is observed in $dN/dy$ and $\langle p_T \rangle$ when we compare the results from isobar collisions with Au + Au and U + U collisions.

## I. INTRODUCTION

The theoretical predictions from quantum chromodynamics (QCD) suggest a transition at sufficiently high temperature and energy density from hadron gas to a deconfined state of quarks and gluons called quark-gluon plasma (QGP) [1–3]. Many experimental evidences of a medium dominated by the partonic degrees of freedom have been reported, which motivates the study of the QGP [4–10]. Bulk observables like collective flow and transverse momentum spectra of produced particles are important probes of the QGP medium. Collective flow is quantified by the coefficients in the Fourier expansion of azimuthal angle distribution of produced particles with respect to the reaction plane, namely, directed flow ($v_1$), elliptic flow ($v_2$), triangular flow ($v_3$), and higher-order harmonics [11,12]. Transverse momentum spectra of produced particles provide information on the transverse expansion of the QGP medium, which is coherently linked to the initial density of the colliding system. The $\langle p_T \rangle$ of particles in heavy-ion collisions is related to the temperature and collective transverse velocity of the medium. A smaller initial transverse size of the collision region would result in a larger entropy density, hence larger temperature which gives rise to an increased radial flow and, consequently, higher $\langle p_T \rangle$ values [13,14]. Various nuclei like Cu, Au, Pb, and U with different shapes and deformation have collided in multiple high-energy colliders to investigate the QGP medium. The different structural properties modify the geometry of the initial density distributions, and therefore such collisions can also be used to characterize the nuclear structure.

In the year 2018, isobar collisions of ${}^{96}_{44}$Ru + ${}^{96}_{44}$Ru and ${}^{96}_{40}$Zr + ${}^{96}_{40}$Zr have been performed at the BNL Relativistic Heavy-Ion Collider (RHIC). The two isobars have the same atomic mass number but different numbers of protons and neutrons. It is suggested that Ru nuclei have quadrupole deformation while the Zr nuclei have octupole deformation [15,16]. A detailed study of the nuclear structures is possible using the available high-statistics data from the experiments. Recent results from the STAR experiment in isobar collisions at $\sqrt{s_{NN}} = 200$ GeV showed a deviation in the $v_2$ and $v_3$ ratios between the two isobars which are attributed to their different nuclear density and deformation [17]. Many studies using a multiphase transport (AMPT) model with different Woods-Saxon (WS) parametrization for isobars have also demonstrated the effects of deformation and neutron thickness on the $v_2$, $v_3$, $\langle p_T \rangle$ and their correlations [15,18–25]. There have been studies using density-functional theory (DFT) to demonstrate the effects of deformation in isobar collisions [26,27]. Hydrodynamic-based models such as Trajectum have also been used to infer nuclear structure in isobar collisions [28]. A recent study has also demonstrated isobar collisions as the ideal test for the baryon junction hypothesis [29].

In this paper, we report a study of $p_T$ spectra for identified particles ($\pi^\pm$, $K^\pm$, $p$, and $\bar{p}$) at midrapidity ($|y| < 0.5$) in isobar collisions (Ru + Ru and Zr + Zr) at $\sqrt{s_{NN}} = 200$ GeV using the AMPT model. We have varied WS parameters to generate different sets of events for isobar collisions and studied effects on the $p_T$ spectra, $dN/dy$, $\langle p_T \rangle$, and particle ratios. The choice of the WS parameters is motivated by the previous studies done with the AMPT model [15,18–21,30]. Additionally, we also studied U + U and Au + Au collisions at $\sqrt{s_{NN}} = 200$ GeV using the AMPT model in order to understand the system size evolution of particle production.

*Corresponding author: vipul.bairathi@gmail.com

The paper is organized in the following order: Section II A describes the AMPT model in brief and various parametrizations of the Woods-Saxon distribution. The section also involves the analysis details. In Sec. III, we present the results on the identified hadron $p_T$ spectra, integrated yield, average transverse momentum, and particle ratios in isobar collisions at $\sqrt{s_{NN}} = 200$ GeV. We present the ratio of the observables between the two isobar systems for various cases of WS parametrization. We also present results from Au + Au and U + U collisions and compare them with the isobar collisions to investigate system size evolution. In Sec. IV, we summarize and discuss the results presented in this paper.

## II. MODELING NUCLEAR STRUCTURE WITH A MULTIPHASE TRANSPORT MODEL

### A. A multiphase transport model

AMPT is a hybrid Monte Carlo event generator extensively used to study relativistic heavy-ion collisions [31]. It has four main components, which include the HIJING [32] model for the initial condition and Zhang's parton cascade (ZPC) [33] for the evolution of the partonic stage. Quark coalescence and Lund string fragmentation models are used to produce hadrons. A relativistic transport (ART) model [34] is used for the final-state hadronic interactions. We have used AMPT string melting model version 2.26t9 with a partonic cross-section of 3 mb to simulate the collision events. The nucleon distribution of nuclei in AMPT is modeled using the WS function defined as follows:

$$\rho(r,\theta) = \frac{\rho_0}{1 + e^{[\{r - R(\theta,\phi)\}/a]}}, \tag{1}$$

where $\rho_0$ is the normal nuclear density, $r$ is the distance from the center of the nucleus, $a$ is the surface diffuseness parameter, and $R(\theta,\phi)$ is the parameter characterizing the deformation of the nucleus,

$$R(\theta,\phi) = R_0[1 + \beta_2 Y_{2,0}(\theta,\phi) + \beta_3 Y_{3,0}(\theta,\phi)]. \tag{2}$$

$R_0$ represents the radius parameter, $\beta_2$ and $\beta_3$ are the quadrupole and octupole deformities, and $Y_{l,m}(\theta,\phi)$ are the spherical harmonics. We studied three different cases of WS parameters for Ru + Ru and Zr + Zr collisions at $\sqrt{s_{NN}} =$ 200 GeV, as shown in Table I [30]. The first case considered is without deformation and has the same values of $a$ and $R_0$ for both nuclei (case 1). In the second case, we consider different values of $a$ and $R_0$ (case 2). In the third case, we also modified the deformation parameters, $\beta_2$ and $\beta_3$, along with the radius and surface diffuseness parameter (case 3). We have also studied two deformation cases for the U nuclei, as shown in Table II [31,35]. A total of nine million minimum bias events for Ru + Ru and Zr + Zr collisions at $\sqrt{s_{NN}} = 200$ GeV have been analyzed for each of the three cases. Nearly two million minimum bias events have been studied for Au + Au and U + U collisions at $\sqrt{s_{NN}} = 200$ GeV.

TABLE I. Various deformation configurations for the Ru and Zr nuclei in the AMPT model

| | Ru | | | Zr | | |
|---|---|---|---|---|---|---|
| Parameter | case 1 | case 2 | case 3 | case 1 | case 2 | case 3 |
| $R_0$ | 5.096 | 5.067 | 5.090 | 5.096 | 4.965 | 5.090 |
| $a$ | 0.540 | 0.500 | 0.460 | 0.540 | 0.556 | 0.520 |
| $\beta_2$ | 0 | 0 | 0.162 | 0 | 0 | 0.060 |
| $\beta_3$ | 0 | 0 | 0 | 0 | 0 | 0.200 |

### B. Analysis details

We have calculated $p_T$ spectra of identified hadrons using the AMPT model. The study is carried out in various centrality classes. The centrality of an event is based on the total multiplicity of charged hadrons produced in the pseudorapidity range $|\eta| < 0.5$ called reference multiplicity. The left panel of Fig. 1 shows reference multiplicity distribution in Ru + Ru collisions at $\sqrt{s_{NN}} = 200$ GeV. Various centrality classes 0%–10%, 10%–20%, 20%–40%, 40%–60%, and 60%–80% are shown in alternative gray and white bands. Similar centrality selection criteria based on multiplicity in $|\eta| < 0.5$ is used for all the other data set. A comparison of the reference multiplicity distribution for various collision systems is shown in the right panel of Fig. 1. Larger multiplicity values are observed in U + U collisions, owing to the highest number of nucleons among all the nuclei studied. We compute the ratio of reference multiplicity between the two isobar systems case-by-case. The left panel of Fig. 2 shows the reference multiplicity ratio between Ru + Ru and Zr + Zr collisions for the three cases of WS parameters from the AMPT model and compared with the STAR experimental data [17]. The reference multiplicity ratio obtained from case 1 is around unity and does not agree with the experimental data. This implies the need for different nuclear structures for isobar nuclei to explain the experimental data. The ratio in case 3 deviates from the experimental data at higher multiplicity. Case 2 data set of the AMPT model seems to better describe the STAR experimental data. Hence, the reference multiplicity ratio shows the influence of the nuclear size and thickness variation [36]. The larger values of reference multiplicity ratio at high multiplicity can be explained as the zirconium nucleus has a smaller effective radius than ruthenium, which leads to a larger probability of high multiplicity events in central Zr + Zr than Ru + Ru collisions [24].

TABLE II. Various deformation configurations for the U and Au nuclei in the AMPT model.

| | U | | Au |
|---|---|---|---|
| Parameter | case 1 | case 2 | |
| $R_0$ | 7.115 | 6.810 | 6.380 |
| $a$ | 0.540 | 0.550 | 0.535 |
| $\beta_2$ | 0 | 0.280 | 0 |
| $\beta_3$ | 0 | 0 | 0 |

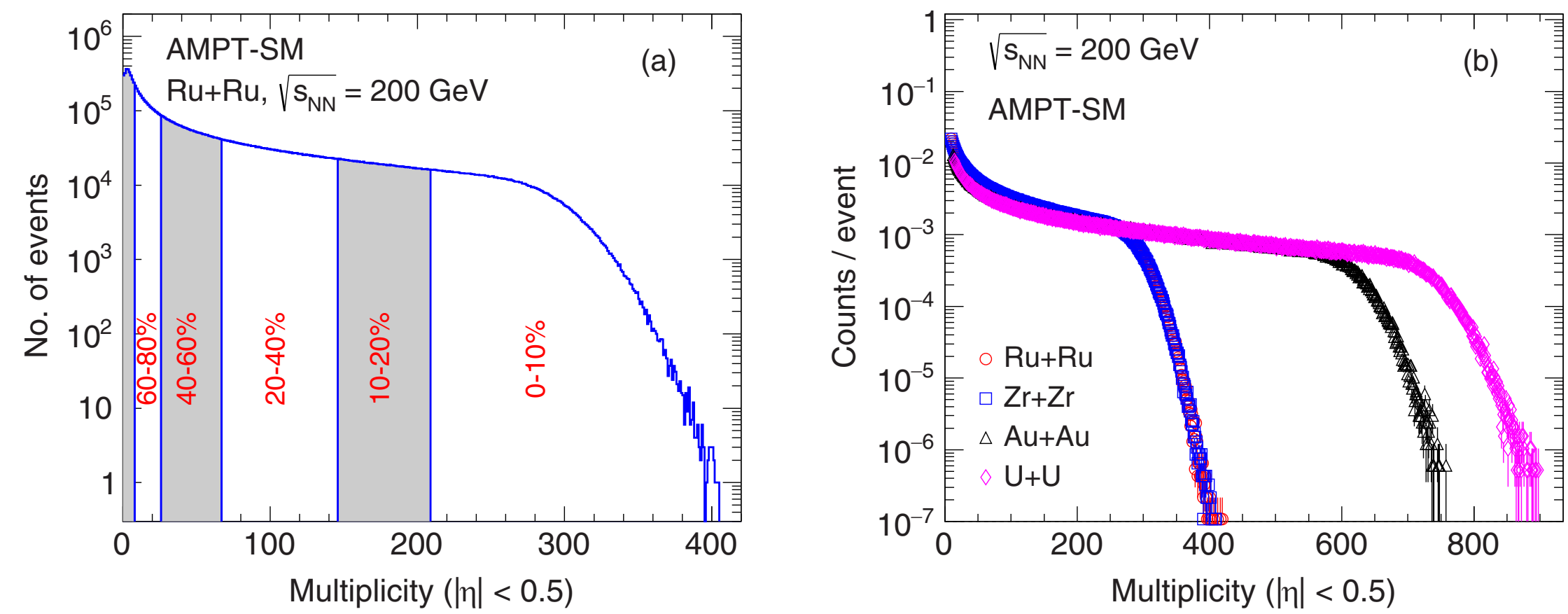

FIG. 1. (a) Multiplicity distribution in $|\eta| < 0.5$ for Ru + Ru collisions at $\sqrt{s_{NN}} = 200$ GeV. Different centrality classes from 0%–10% to 60%–80% are shown in alternate gray and white bands. (b) Multiplicity distributions in $|\eta| < 0.5$ for Ru + Ru, Zr + Zr, Au + Au, and U + U collisions at $\sqrt{s_{NN}} = 200$ GeV.

Elliptic flow, which primarily arises due to the initial spatial anisotropy of the overlap region in heavy-ion collisions, is also studied for the three configurations of WS parameters of the isobar nuclei and compared with the results from the STAR experiment [17]. The $p_T$-integrated elliptic flow of charged hadrons is evaluated using the equation $v_2 = \langle\cos(2\phi - 2\Psi_2)\rangle$, where $\Psi_2$ is the second-order event plane angle [11]. The ratio of $p_T$-integrated $v_2$ between Ru + Ru and Zr + Zr collisions at $\sqrt{s_{NN}} = 200$ GeV for charged hadrons as a function of centrality case-by-case are shown in the right panel of Fig. 2 and compared with the STAR data. The $v_2$ ratio from case 1 is close to unity and significantly deviates from the experimental data. In central collisions, the $v_2$ ratio is explained by case 3, whereas case 2 deviates from the data. This shows the dominance of deformation over nuclear skin effect on $v_2$ in central isobar collisions. Both cases 2 and 3 describe the data in peripheral collisions within statistical uncertainties. This implies that $v_2$ ratio in peripheral collisions are primarily affected by the nuclear skin [20].

We calculate $p_T$ spectra for $\pi^\pm$, $K^\pm$, $p$, and $\bar{p}$ at midrapidity in various centrality classes in different collision systems at $\sqrt{s_{NN}} = 200$ GeV. We compute $dN/dy$ and $\langle p_T\rangle$ for identified hadrons by integrating $p_T$ spectra over a measured $p_T$ range for each centrality class. The magnitude and trend of $\langle p_T\rangle$ do not change significantly if we integrate over a larger $p_T$ range. We report ratios of $dN/dy$ and $\langle p_T\rangle$ between different systems with various deformation parameters to observe the effect of nuclear structure on particle production in relativistic heavy-ion collisions.

## III. RESULTS

In this section, we present results on transverse momentum spectra of identified hadrons ($\pi^\pm$, $K^\pm$, $p$, and $\bar{p}$) at midrapidity ($|y| < 0.5$) in isobar (Ru + Ru and Zr + Zr), Au + Au, and U + U collisions at $\sqrt{s_{NN}} = 200$ GeV using the AMPT model. Particle yield and average transverse momentum extracted from the $p_T$ spectra are also presented. Ratios of these quantities among the systems are discussed to investigate the dependence of nuclear structure and system size on particle production in heavy-ion collisions.

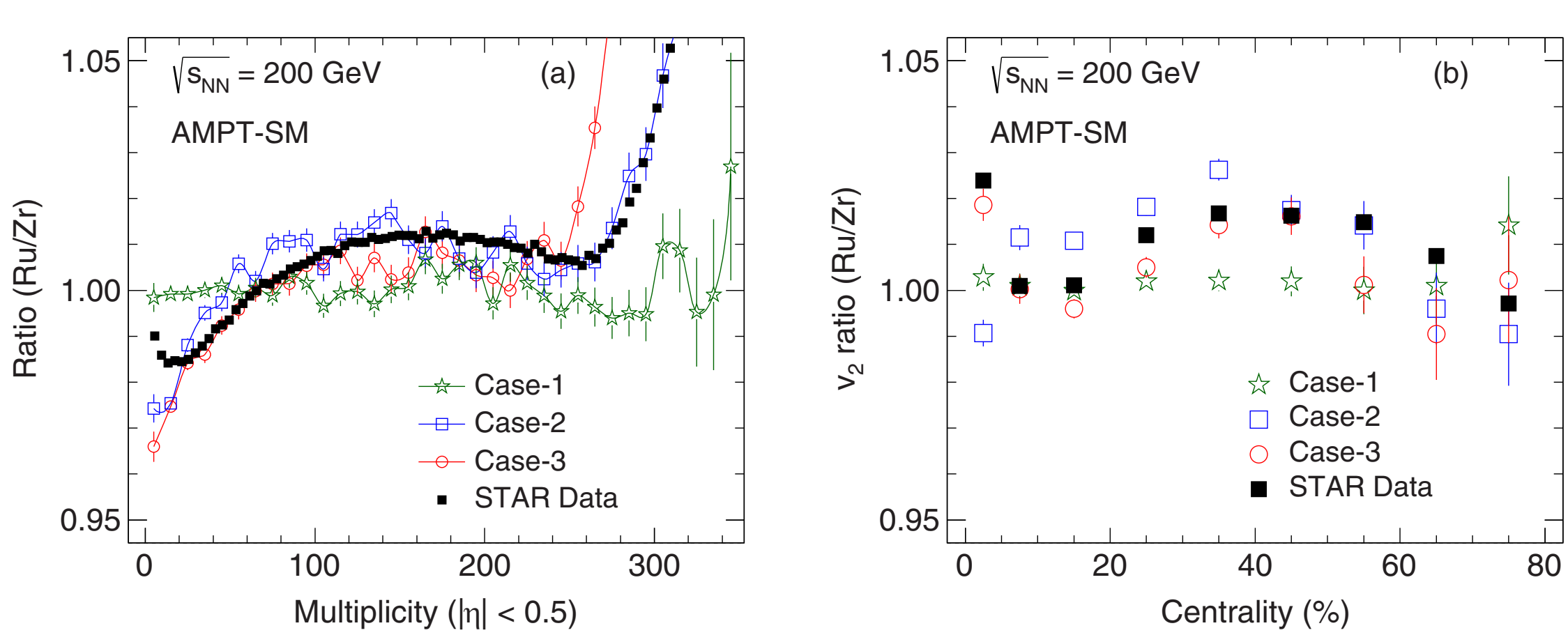

FIG. 2. (a) Multiplicity ratio, (b) $v_2$ ratio between Ru + Ru and Zr + Zr collisions at $\sqrt{s_{NN}} = 200$ GeV for three different WS parametrization compared with the STAR data [17].

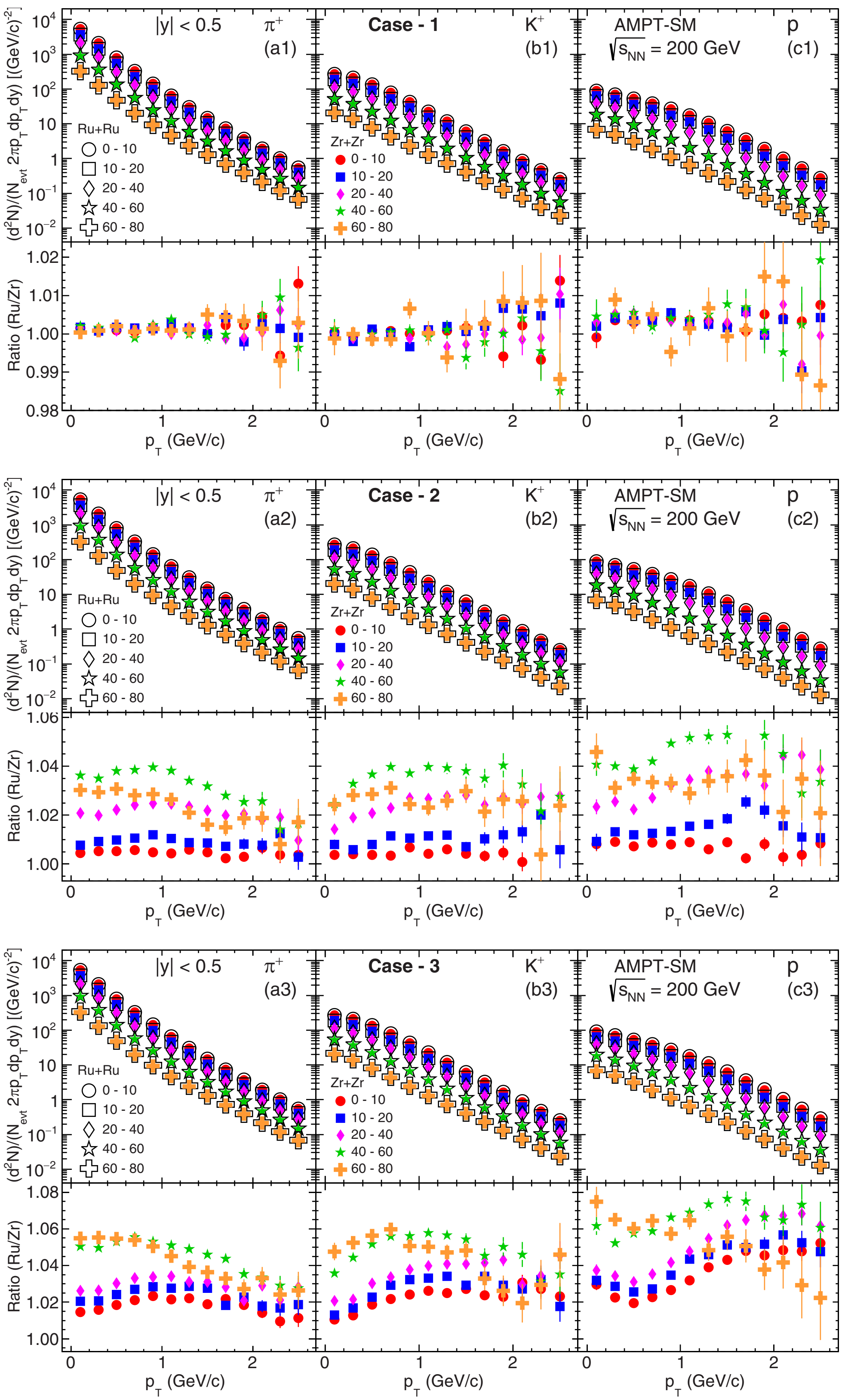

FIG. 3. $p_T$ spectra for $\pi^+$, $K^+$, and $p$ at midrapidity in Ru + Ru and Zr + Zr collisions and the corresponding ratio between the two systems at $\sqrt{s_{NN}} = 200$ GeV with three different cases of WS parametrization.

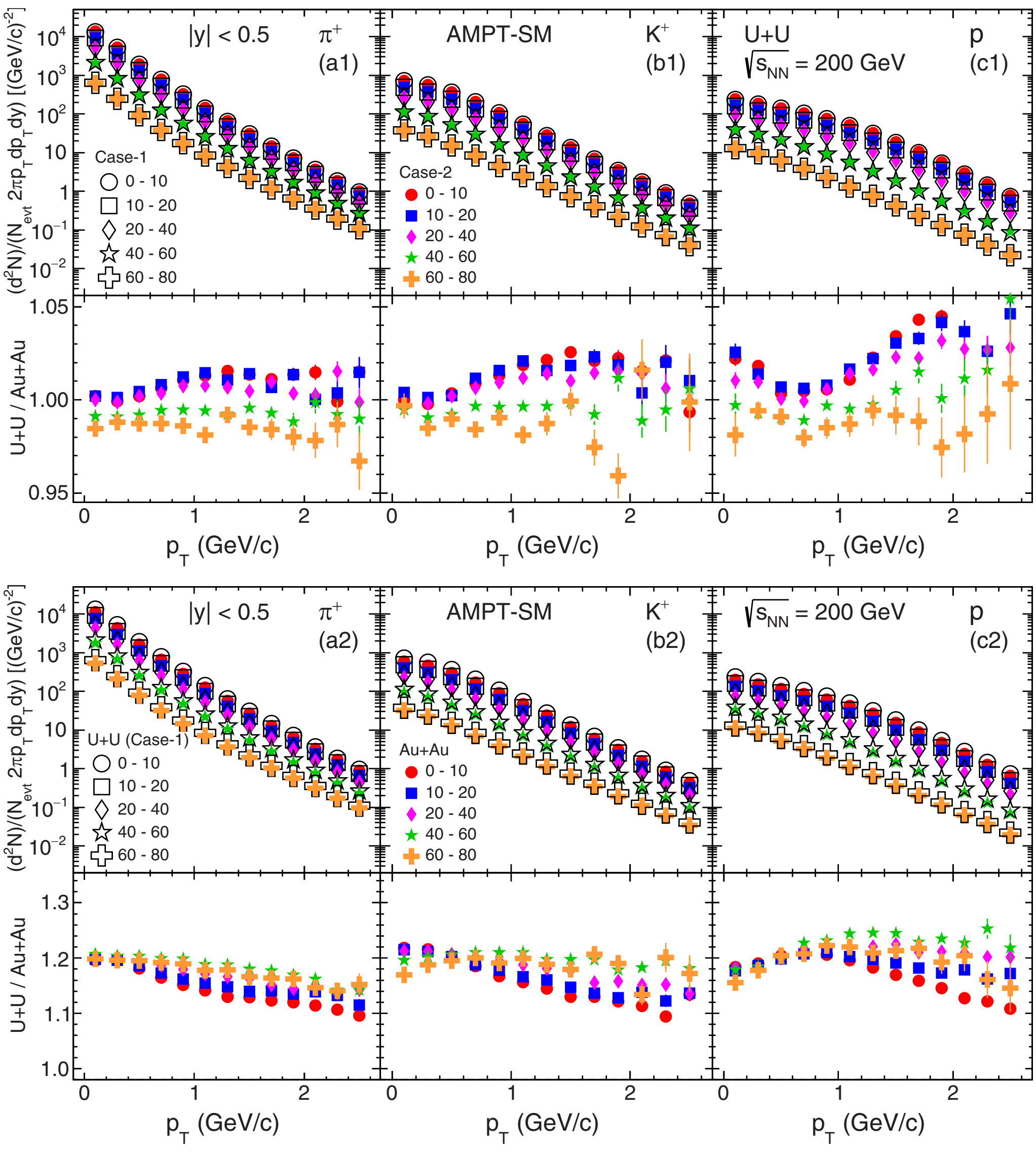

FIG. 4. $p_T$ spectra for $\pi^+$, $K^+$, and $p$ at midrapidity in U + U collisions with two cases of WS parametrization and Au + Au collisions at $\sqrt{s_{NN}} = 200$ GeV. The bottom panels show the ratio between the two cases in U + U collisions and between U + U (case 1) and Au + Au collisions.

### A. Transverse momentum ($p_T$) spectra

Transverse momentum spectra of $\pi^+$, $K^+$, and $p$ at midrapidity ($|y| < 0.5$) in Ru + Ru and Zr + Zr collisions at $\sqrt{s_{NN}} = 200$ GeV for three cases of WS parametrization are shown in Fig. 3. We have calculated $p_T$ spectra in five different centrality classes from 0%–10% to 60%–80%. We observed a clear centrality dependence in the $p_T$ spectra of all the particle species. The slopes of the $p_T$ spectra get flattened as centrality changes from peripheral to central, indicating the effect of stronger radial flow in central collisions. The slope of $p_T$ spectra for protons is much flatter than the $p_T$ spectra of pions and kaons, which is also consistent with an increase in radial flow with increasing particle mass. These observations are similar in all three cases of WS parametrization of isobar collisions. The bottom panels in each case show the ratio of $p_T$ spectra between the Ru + Ru and Zr + Zr collisions. We do not observe any deviation from unity in case 1 within the statistical uncertainties. The ratio of $p_T$ spectra in case 2 shows a deviation from unity. The deviation increases from central to peripheral collisions. Pions and kaons show a maximum deviation of 4% while protons show 6% deviation. In case 3, when we include a difference in deformation along with the nuclear skin and size between the isobar nuclei, we observe even higher deviation from unity reaching up to 6% for pions and kaons and 8% for protons in the measured $p_T$ region.

We also show the identified hadron $p_T$ spectra for two cases of WS parametrization as mentioned in Table II to study the effect of nuclear structure in uranium nuclei. Figure 4 shows the $p_{\rm T}$ spectra and their ratio in collisions of nondeformed and deformed U nuclei at $\sqrt{s_{\rm NN}} = 200$ GeV. Figure 4 also shows the $p_{\rm T}$ spectra of identified hadrons in Au + Au collisions at $\sqrt{s_{\rm NN}} = 200$ GeV and their ratio with U + U (case 1) collisions. We observe a similar centrality dependence of the $p_{\rm T}$ spectra in Au + Au and U + U collisions as in the case of isobar collisions. The similar radial flow effect on the slope of $p_{\rm T}$ spectra is also observed. We observe nearly 2% deviation from unity in the ratio of $p_{\rm T}$ spectra between the two cases of uranium nuclei. The ratio is higher than unity for central collisions, whereas, lower in peripheral collisions due to the structural differences between the two cases. The spectra ratio shows 10%–20% higher invariant yield in U + U collisions than Au + Au collisions. This is due to a larger number of participant nucleons in U + U than Au + Au collisions. We observe a centrality dependence at higher $p_{\rm T}$ in the spectra ratio.

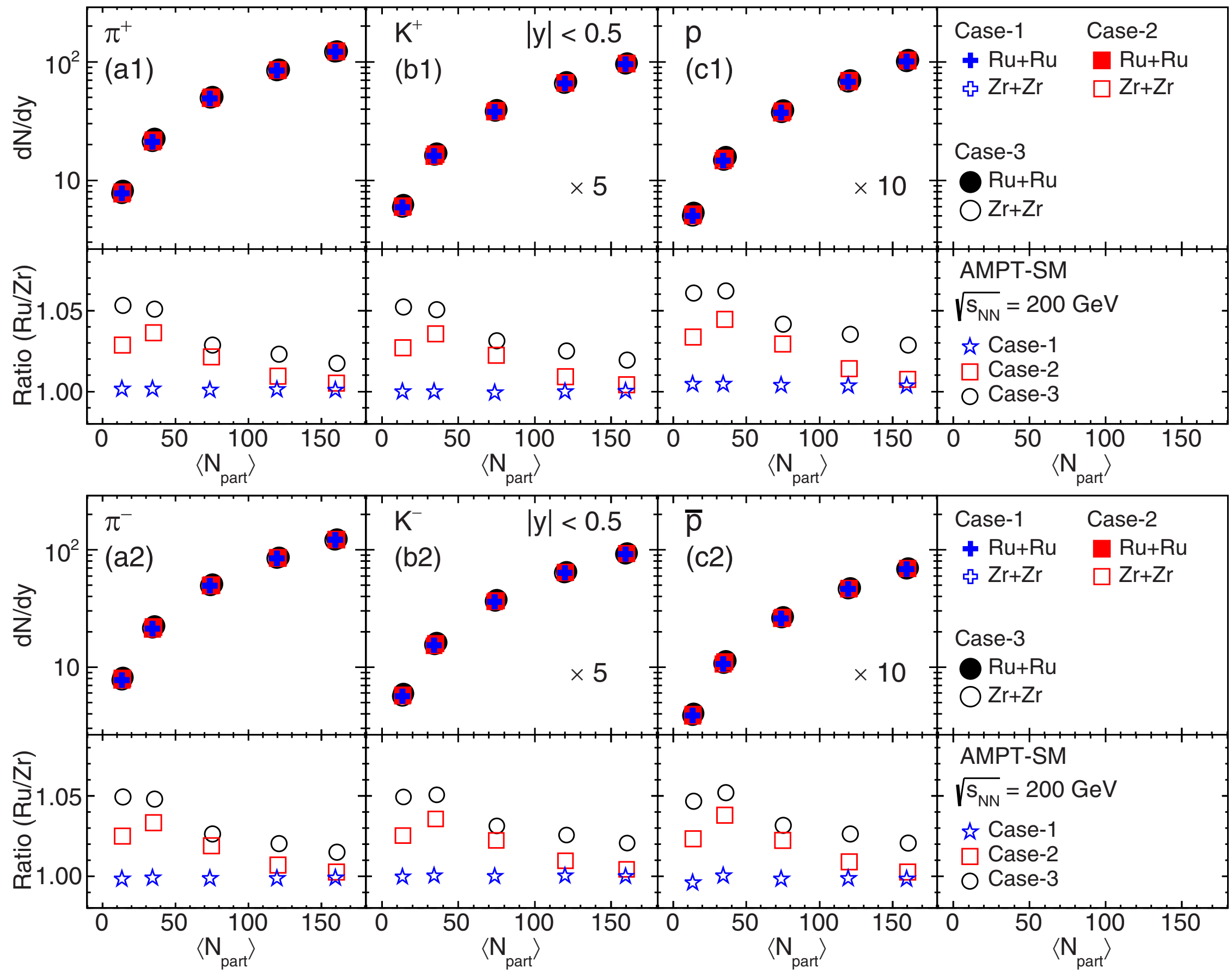

FIG. 5. Particle yield ($dN/dy$) of $\pi^{\pm}$, $K^{\pm}$, $p$, and $\bar{p}$ as a function of $\langle N_{\rm part}\rangle$ in Ru + Ru and Zr + Zr collisions at $\sqrt{s_{\rm NN}} = 200$ GeV for three WS parametrization. $dN/dy$ of kaons and (anti-)protons are scaled for better representation with factors of five and ten, respectively. The ratio of $dN/dy$ between the two systems is shown in the bottom panels.

### B. Particle yield and average transverse momentum

The $p_{\rm T}$-integrated yield ($dN/dy$) of pions, kaons, and (anti-)protons at midrapidity ($|y| < 0.5$) as a function of the average number of participating nucleons ($\langle N_{\rm part}\rangle$) in Ru + Ru and Zr + Zr collisions at $\sqrt{s_{\rm NN}} = 200$ GeV from the AMPT model for three cases of WS parameters is shown in Fig. 5. $dN/dy$ increases from peripheral to central collisions for all the particles studied. The three WS parametrizations have similar increasing trends with $\langle N_{\rm part}\rangle$.

The ratio of $dN/dy$ between the two systems is also shown in the bottom panels of Fig. 5. The ratio shows no deviation from unity in case 1 of isobar nuclei with the same nuclear size and without deformation. We observe a deviation from unity in the ratio of particle yields between the two nuclei for case 2, which arises due to the nuclear size and skin effect [37]. In case 3, we observe a more significant deviation in the ratio of particle yields which could be attributed to the inclusion of deformation along with different nuclear sizes. A clear centrality dependence of the ratio is observed with a maximum deviation of 5% from unity in peripheral collisions compared with the central collisions.

Average transverse momentum quantifies the shape of $p_{\rm T}$ spectra. Figure 6 shows the evolution of $\langle p_{\rm T}\rangle$ of pions, kaons, and (anti-)protons at midrapidity in Ru + Ru and Zr + Zr collisions at $\sqrt{s_{\rm NN}} = 200$ GeV as a function of $\langle N_{\rm part}\rangle$ from the AMPT model for three cases of WS parameters. The $\langle p_{\rm T}\rangle$ of pions and kaons show weak dependence on $\langle N_{\rm part}\rangle$ for all the three cases of WS parametrization, while protons and antiprotons show an increase from peripheral to central collisions. Calculations based on various hydrodynamic models reported a clear decreasing trend of $\langle p_T \rangle$ from higher $N_{\text{part}}$ to lower $N_{\text{part}}$ for all charged particles [21,28]. The AMPT model does not describe the centrality dependence of the $\langle p_T \rangle$ for pions and kaons in isobar collisions with a partonic cross-section of 3 mb. This was also pointed out in AMPT simulations of Pb + Pb and U + U collisions [38,39]. The bottom panels of Fig. 6 show the $\langle p_T \rangle$ ratio between Ru + Ru and Zr + Zr collisions. The ratio of $\langle p_T \rangle$ for particles and antiparticles shows no deviation from unity for the isobar nuclei having the same nuclear size and no deformation. We observe a deviation from unity within ±1% in nuclei with different nuclear sizes and deformations. The ratio for (anti-)protons shows a large deviation compared with pions and kaons. The deviation seems to increase with particle mass, which could result from increasing radial flow in central collisions.

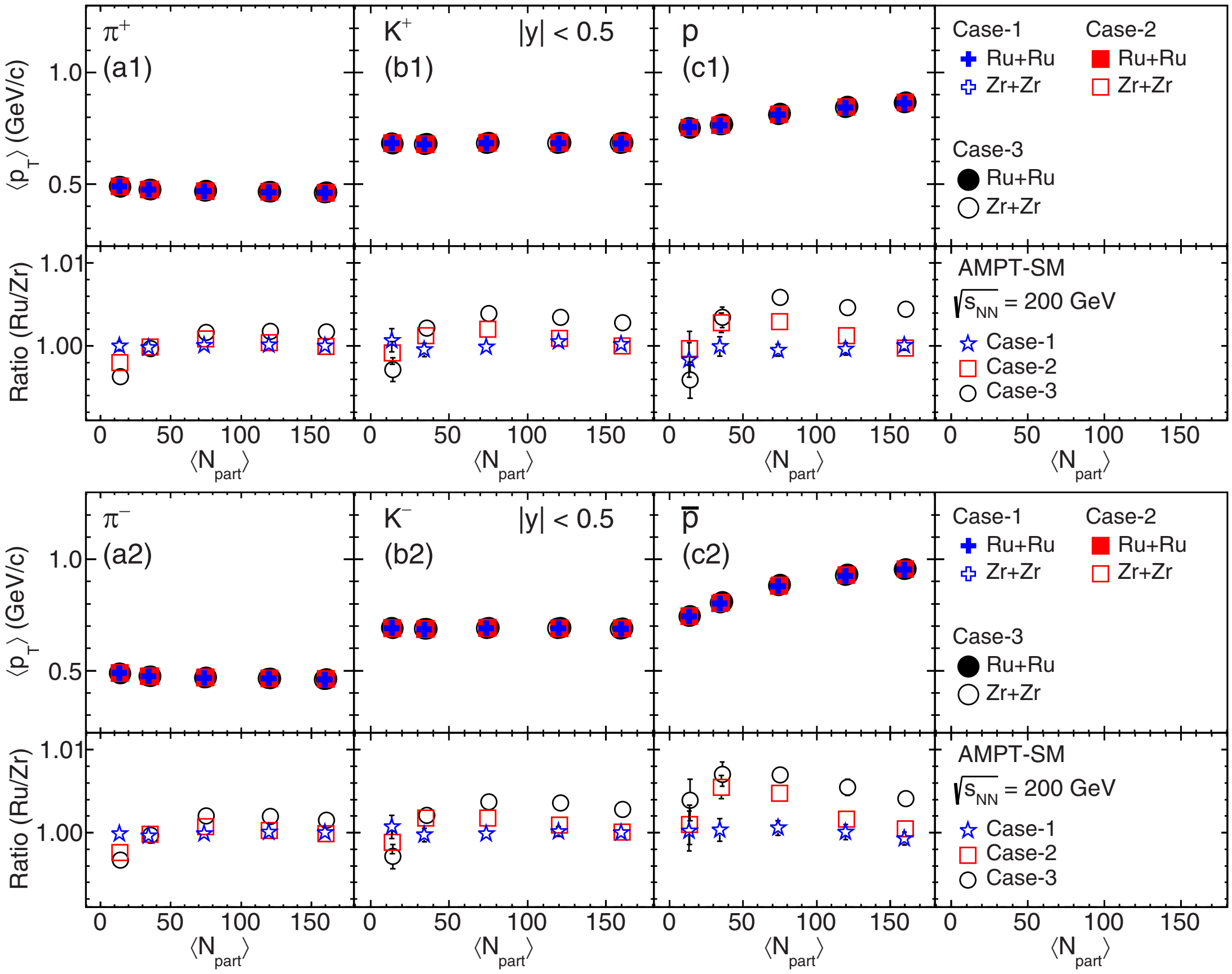

FIG. 6. Average transverse momentum ($\langle p_T \rangle$) of $\pi^{\pm}$, $K^{\pm}$, $p$, and $\bar{p}$ as a function of $\langle N_{\text{part}} \rangle$ in Ru + Ru and Zr + Zr collisions at $\sqrt{s_{\text{NN}}} = 200$ GeV for three WS parametrization. The ratio of $\langle p_T \rangle$ between the two systems is shown in the bottom panels.

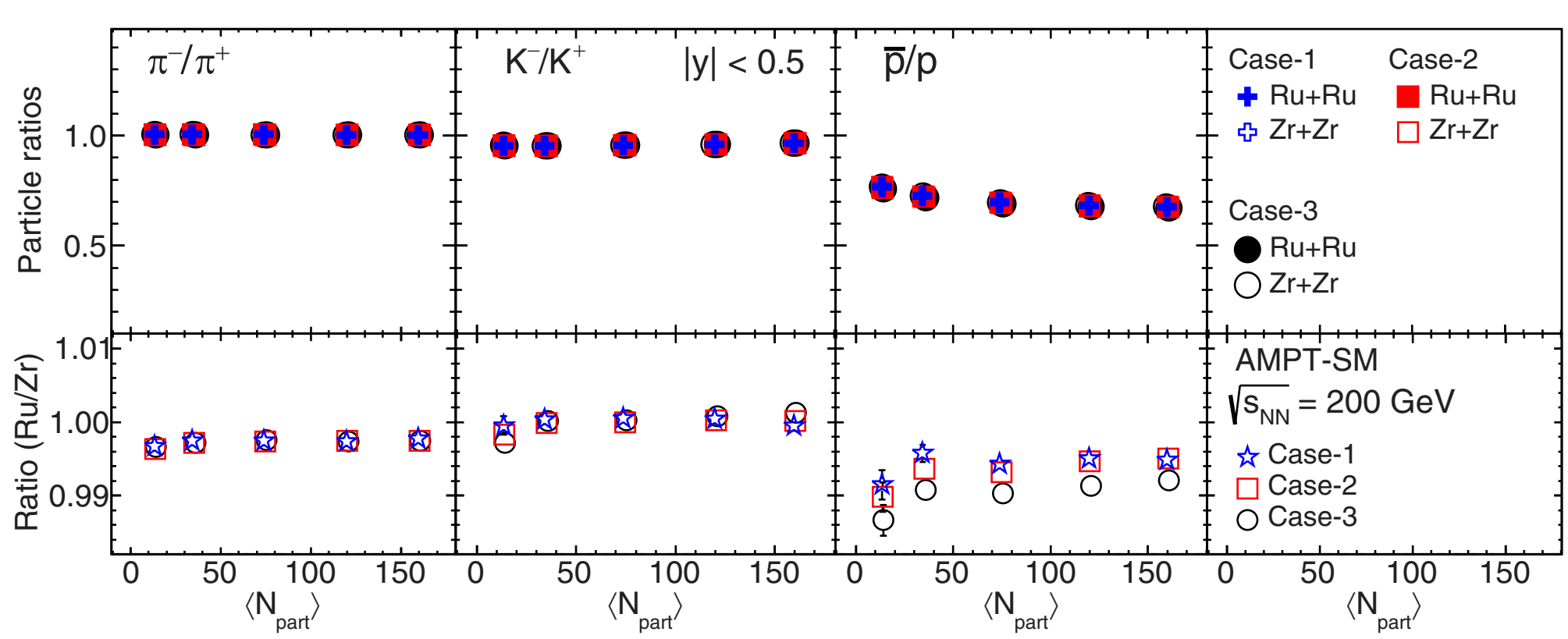

FIG. 7. Particle ratios ($\pi^-/\pi^+$, $K^-/K^+$, and $\bar{p}/p$) as a function of $\langle N_{\text{part}} \rangle$ in Ru + Ru and Zr + Zr collisions at $\sqrt{s_{\text{NN}}} = 200$ GeV for three cases of WS parametrization. The ratio between the two systems is shown in the bottom panels.

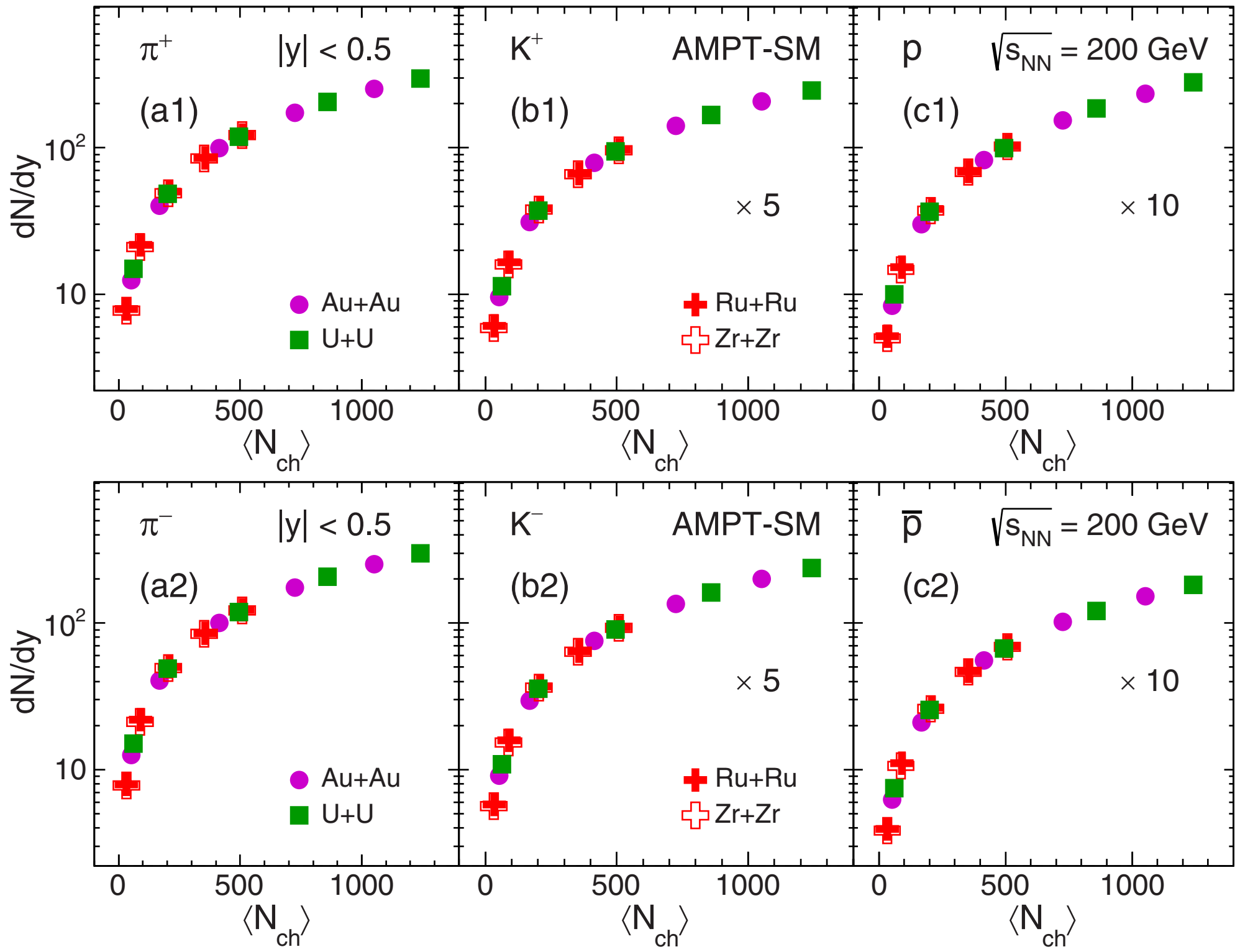

FIG. 8. $dN/dy$ of $\pi^{\pm}$, $K^{\pm}$, $p$, and $\bar{p}$ as a function of $\langle N_{ch}\rangle$ at midrapidity in Ru + Ru, Zr + Zr, Au + Au, and U + U collisions at $\sqrt{s_{\rm NN}}$ = 200 GeV. The $dN/dy$ of kaons is scaled with a factor of five and protons with a factor of ten for better representation.

### C. Particle ratios

The ratio of particle yields helps us understand the relative particle production in heavy-ion collisions. Figure 7 shows the antiparticle to particle ratios, $\pi^-/\pi^+$, $K^-/K^+$, and $\bar{p}/p$ as a function of $\langle N_{\rm part}\rangle$ in Ru + Ru and Zr + Zr collisions at $\sqrt{s_{\rm NN}}$ = 200 GeV for the three WS parametrization. The ratio $\pi^-/\pi^+$ and $K^-/K^+$ are close to unity, while $\bar{p}/p$ ratio is lower than unity by ≈20%–30% in both the collision systems. The $\bar{p}/p$ ratio decreases slightly from peripheral to central collisions for all three cases of WS parametrization. It is interesting to observe that the ratios in the three cases of WS parametrization are similar because of the cancellation of nuclear geometry effect. There is a difference in the proton and neutron number between the incoming Ru nucleus ($Z = 44$) and Zr nucleus ($Z = 40$) which may cause these particle ratios to be different between the two colliding systems. Therefore, we studied the ratio of antiparticle to particle ratios between Ru + Ru and Zr + Zr collisions as shown in the bottom panels of Fig. 7. We observed that the ratio of $\pi^-/\pi^+$ between the two systems is lower than unity. The difference in the number of $u$ and $d$ quarks is higher in the Zr nucleus than in the Ru nucleus. This effect of isospin causes a higher $\pi^-/\pi^+$ ratio in Zr + Zr collisions compared with Ru + Ru collisions. The ratio of $\bar{p}/p$ is further lower than the ratio of $\pi^-/\pi^+$ due to the additional baryon stopping process along with the isospin effect. The isospin asymmetry is expected to be increasingly important in isobar collisions which result in higher protons in Ru + Ru collisions and, consequently, more baryon stopping than in Zr + Zr collisions [40–42]. In the $K^-/K^+$ ratio, the magnitude is close to unity, indicating the dominance of the pair production mechanism in producing kaons.

### D. System size dependence

Figure 8 shows particle yields of $\pi^{\pm}$, $K^{\pm}$, $p$, and $\bar{p}$ as a function of average charged particle multiplicity ($\langle N_{ch}\rangle$) in various collision systems (Ru + Ru, Zr + Zr, Au + Au, and U + U) at $\sqrt{s_{\rm NN}}$ = 200 GeV. The $\langle N_{ch}\rangle$ is the total yield of pions, kaons, and protons calculated in midrapidity ($|\eta| < 1.0$). The final-state average charged particle multiplicity reaches a higher magnitude in U + U collisions than the isobar collisions due to more number of participating nucleons and energy density. An increase in $dN/dy$ is observed with increasing $\langle N_{ch}\rangle$ for all the particle species in all collision systems. Particle yields for different colliding systems show a smooth variation with $\langle N_{ch}\rangle$. Figure 9 shows average transverse momentum of $\pi^{\pm}$, $K^{\pm}$, $p$, and $\bar{p}$ as a function of $\langle N_{ch}\rangle$ at midrapidity in Ru + Ru and Zr + Zr collisions compared with Au + Au and U + U collisions at $\sqrt{s_{\rm NN}}$ = 200 GeV. The magnitude of $\langle p_{\rm T}\rangle$ increases with increasing particle mass which can be attributed to the stronger radial flow. The $\langle p_{\rm T}\rangle$ also shows a smooth variation with $\langle N_{ch}\rangle$ for all the (anti-)particles independent of colliding systems. The $\langle p_{\rm T}\rangle$ of $\pi^{\pm}$, $K^{\pm}$ shows weak dependence with $\langle N_{ch}\rangle$ whereas $\langle p_{\rm T}\rangle$ of $p(\bar{p})$ increases with $\langle N_{ch}\rangle$.

## IV. SUMMARY AND DISCUSSIONS

We present predictions of the transverse momentum spectra for $\pi^{\pm}$, $K^{\pm}$, $p$, and $\bar{p}$ in ${}^{96}_{44}$Ru + ${}^{96}_{44}$Ru and ${}^{96}_{40}$Zr + ${}^{96}_{40}$Zr collisions at $\sqrt{s_{NN}} = 200$ GeV using the AMPT model. We expect a similar medium to be formed in the isobar collisions, given that these nuclei have the same mass number implying an equal number of participating nucleons and binary collisions. Any observed difference in the $p_T$ spectra, $dN/dy$, and $\langle p_T \rangle$ may indicate a variation in nuclear size and structure between the two isobar nuclei. We study the effect of nuclear size and deformation on the $p_T$ spectra, $dN/dy$, and $\langle p_T \rangle$. The ratios of $dN/dy$ and $\langle p_T \rangle$ between the two isobars show no deviation from unity for the case with no deformation and the same nuclear size. A difference in $dN/dy$ and $\langle p_T \rangle$ is observed between the isobar collisions when we incorporate a different nuclear size for the two isobar nuclei. The maximum deviation is observed when we include quadrupole and octupole deformation along with the difference in nuclear size between the two isobars. Higher deviation in the ratio of $dN/dy$ and $\langle p_T \rangle$ is observed in the case of baryons (protons) than the mesons (pions and kaons). In conclusion, the effect of nuclear structure on particle production is similar for all particle species.

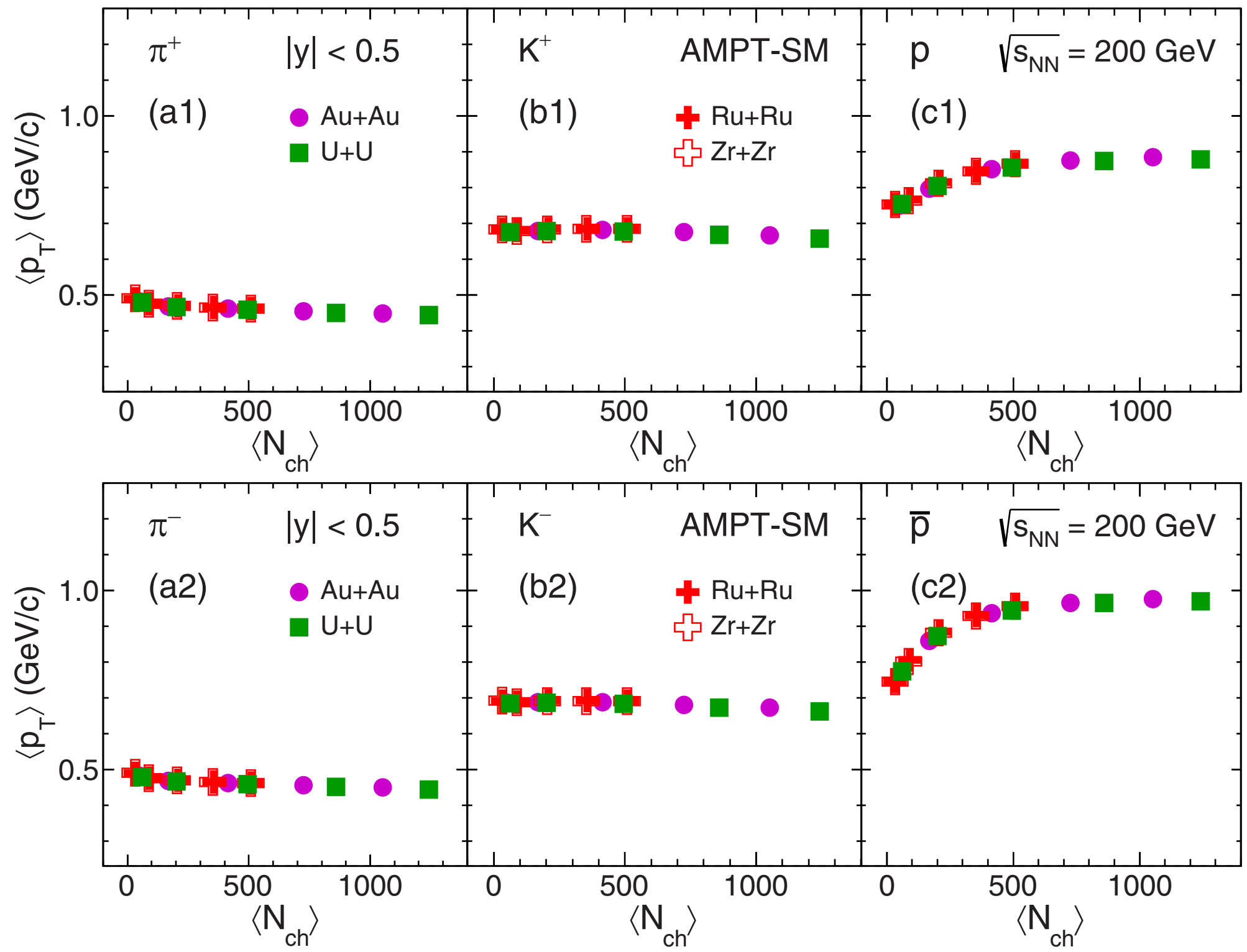

FIG. 9. $\langle p_T \rangle$ of $\pi^{\pm}$, $K^{\pm}$, $p$, and $\bar{p}$ as a function of $\langle N_{ch} \rangle$ at midrapidity in Ru + Ru, Zr + Zr, Au + Au, and U + U collisions at $\sqrt{s_{NN}} =$ 200 GeV.

Relative particle production of identified hadrons is studied in isobar collisions via particle ratios. The ratio of yields of antiparticle to particle cancels the geometrical effect of the nuclei. These ratios between the two isobar systems show a slight deviation from unity for $\pi^{\pm}$ and $p(\bar{p})$ indicating an effect of isospin in the incoming Ru nucleus compared with the Zr nucleus. An additional baryon stopping effect is observed for $\bar{p}/p$ ratio. The $K^-/K^+$ ratio between these isobar collisions is close to unity due to the dominance of the pair production mechanism in kaon production. We also studied the $p_T$ spectra in Au + Au and U + U collisions at $\sqrt{s_{NN}} =$ 200 GeV. $dN/dy$ as a function of $\langle N_{ch} \rangle$ shows an increase from isobar collisions to U + U collisions. $dN/dy$ and $\langle p_T \rangle$ varies smoothly with collision systems for all particles and antiparticles. This shows the system size dependence of the particle production in heavy-ion collisions. This study would help to determine the best nuclear structure parameters to explain the isobar data from the STAR experiment at RHIC.

## ACKNOWLEDGMENTS

C.J. acknowledges the financial support from DAE-DST, Government of India bearing Project No. SR/MF/PS-02/2021-IISERT (E-37130) and S.K. acknowledges partial financial support received from the Agencia Nacional de Investigación y Desarrollo (ANID), Chile, from the ANID FONDECYT regular 1230987 Etapa 2023, Chile, and from the ANID PIA/APOYO AFB220004, Chile.